\documentstyle[12pt]{article}
\font \tbfontt                = cmbx12
\font \tbfonts                = cmbx7  at 8.4 pt
\font \tbfontss               = cmbx5  at 6 pt
\newfam\cmbxfam
\textfont\cmbxfam=\tbfontt \scriptfont\cmbxfam=\tbfonts
\scriptscriptfont\cmbxfam=\tbfontss
\def\bf{\fam\cmbxfam\tbfontt}
\begin{document}
\renewcommand{\theequation}{\arabic{section}.\arabic{equation}}
\parskip = 0 pt
\baselineskip = 16 pt
\parindent = 20 pt
\centerline{\LARGE  Time scales in quantum mechanics}\par
\centerline{\LARGE  by a scattering map}\par
\vskip 15  pt
\centerline{L.~Lanz\footnote{Dipartimento di Fisica
dell'Universit\`a di Milano and Istituto Nazionale di Fisica
Nucleare, Sezione di Milano, Via Celoria 16, I-20133, Milan,
Italy. E-mail: lanz@mi.infn.it} and B.~Vacchini\footnote{Dipartimento di Fisica
dell'Universit\`a di Milano and Istituto Nazionale di Fisica
Nucleare, Sezione di Milano, Via Celoria 16, I-20133, Milan,
Italy. E-mail: vacchini@mi.infn.it}}
\vskip 15 pt
\centerline{\sc Abstract}\par
\vskip 15 pt
{
\baselineskip=12pt
Inside quantum mechanics the problem of decoherence 
for an isolated, finite
system is linked to a coarse-grained description of its dynamics.
}\par
\vskip 25 pt
\hrule
\par
\vfill\eject
\par
\section{Introduction}
\par
As it has been pointed out long ago by Ludwig the typical
problems about the foundations of quantum mechanics (QM), basically
arising from non-separability, as the EPR paradox and
macroscopic superpositions in measuring processes, are avoided if
one shifts the basic elements of reality with which the theory is
dealing from microphysical components of matter to the
macroscopic setup of any experiment; then microsystems are derived
objects carrying correlations and interactions between sources
and detectors. The microphysical structure of matter operatively  implies
the existence of this utmost simple interaction
channel between systems and QM is just its
beautifully simple theory, that one can derive  from
axioms superposed on an objective description of  macrosystems
(Ludwig, 1983), arriving in this way to the
modern formulation in terms of POV measures and 
instruments (Kraus, 1983; Holevo, 1982; Davies, 1976). 
So the real challenge of QM
is the objective description of  macrosystems, where we are
giving this name to any part of the world that is separated from
the environment and prepared in such a way that some
objective, and in this sense classical,
description of it can be given.
Adopting this viewpoint has  very relevant consequences on the
mathematical setting of the theory.
Looking at any realistic example of a quantum description of a
 macrosystem there can be no doubt that its preparation
has to be described by a mixture and not by a pure state.
Preparations of the system are elements of the set ${\cal K({\cal
H})}$ of
statistical operators on the Hilbert space ${\cal H}$, which is
naturally located in the Banach space ${\cal T}({\cal H})$
of trace-class  operators
[it is the base of the positive cone and the
whole space
${\cal T}({\cal H})$
is ``positively generated'' by it (Davies, 1976)].
Then the most natural
way of representing the transformation of a preparation, like the
very fundamental one arising by the time evolution of an
isolated system, is by means of a mapping ${\cal M}$  on ${\cal T}({\cal H})$
transforming
${\cal K}$
into ${\cal K}$.
However such a mapping
is not in general an isometry, this happens if and only if it has an
inverse and it is then implemented by a
unitary mapping on ${\cal H}$
($
{\cal M} \cdot =
{\hat U}\cdot
{\hat U}^{\scriptscriptstyle\dagger}
$), otherwise one meets a truly more general
framework for the dynamics, in which irreversibility appears as
the typical new feature.
On the contrary if one takes as starting point  QM of
microsystems, the use of pure states is usually a very
appropriate idealization, justified by the high level of
experimental control by which a few-particle system can be
prepared and strongly supported by the outstanding role of
unitary representations of symmetries in particle physics: in
this context time evolution is described in the most natural
way by the Schr\"odinger equation. Tackling the
question of  macrosystems one comes to  statistical operators
invoking incomplete information about initial state, or
decoherence by the environment, or some mathematical
extrapolation to infinite system. The last point of view gives
nice results, but only in the particular case of systems at
equilibrium. So the usual scenery for the extraordinary
performances of QM is not very satisfactory: no
direct objectivity can be attributed to particles and macrosystems
(not at equilibrium), by which in Bohr's philosophy such
objectivity can be recovered, can be described only in a
thermodynamic limit, that is hardly
compatible with non-equilibrium situations.
\par
\section{Dealing with finite macrosystems}
\par
The very concept of isolated  macrosystem  is slippery: the
 macrosystem  must be separated inside a spatial region $\omega$
by a suitable preparation
procedure covering a finite time interval $[T,t_0]$, that
will be called ``preparation time''. We will not take the
limit
$T\rightarrow -\infty$, since in our opinion one should
avoid shifting this problem to a cosmological level.
Considering
a finite preparation time means that some memory loss
is operatively necessary,
the price of some coarse graining of the dynamical
description must be paid: to do this we associate in a systematic
way to the preparation procedure a suitable
time scale. The relevant role of the preparation procedure means
a breaking of basic space-time symmetry by suitable boundary
conditions which introduce the peculiarities of the system,
hiding the more universal behavior of local or short range
interactions. The field theoretical approach, that is anyway
mandatory in the relativistic case, is best suited to express the
interplay of local universality and peculiar boundary conditions.
In this discussion of  macrosystems let us take, in the
nonrelativistic limit, a very schematic model, built by one type
of molecules
confined inside a region $\omega$
and interacting by a two body potential $V(
\left |
{\mbox{\bf x}}-{\mbox{\bf y}}
\right |
)$; this system is described by a quantum Schr\"odinger field
(QSF) ${\hat \psi}({\mbox{\bf x}})$, to which
the following local Hamiltonian density is associated:
        \begin{equation}
        \label{1.1}
        {\hat e}({\mbox{\bf x}})
        =
        {{  
        \hbar^2  
        \over  
               2m  
        }}
        \nabla
        {\hat \psi}^{\scriptscriptstyle\dagger}({\mbox{\bf x}})  
        \cdot  
        \nabla
        {\hat \psi}({\mbox{\bf x}})
        +  
        {1\over 2}  
        \int_{\omega} d^3\! {\mbox{\bf  y}}
	\,  
        {\hat \psi}^{\scriptscriptstyle\dagger}
        ({\mbox{\bf x}})
                {\hat \psi}^{\scriptscriptstyle\dagger}  
        ({\mbox{\bf y}})
        V(|{\mbox{\bf x}}-{\mbox{\bf y}}|)
        {\hat \psi}  
        ({\mbox{\bf y}})
        {\hat \psi}  
        ({\mbox{\bf x}})
        \end{equation}
        \[
        {  
        \left[  
        {\hat \psi}({\mbox{\bf x}}),
        {\hat \psi}^{\scriptscriptstyle\dagger}({\mbox{\bf x}}')
        \right]  
        }_\pm  
        =  
        \delta ({\mbox{\bf x}}-{\mbox{\bf x}}').
        \] 
Let us consider a complete orthonormal set of eigenfunctions
$u_f\in L^2(\omega)$,
        \begin{equation}
        \label{1.2}
        -{
        \hbar^2  
        \over  
        2m  
        }  
        \Delta_2  
        u_f({\mbox{\bf x}})= W_f
        u_f({\mbox{\bf x}}) \quad {\mbox{\bf x}}\in\omega,
        \qquad  
        u_f({\mbox{\bf x}})=0 \quad {\mbox{\bf x}}\in  
        \partial\omega ,
        \end{equation}
corresponding to the numerable set of eigenvalues $W_f$, and
the confined QSF ${\hat \psi}_{\scriptscriptstyle C}({\mbox{\bf x}})
        =
        \sum_f u_f ({\mbox{\bf x}}) {\hat
        a}_f $.
It will replace ${\hat \psi}({\mbox{\bf x}})$ in
(\ref{1.1}) and we shall leave out the suffix
C. The Hamiltonian ${\hat H}$ and mass operator ${\hat
M}$ will be taken respectively as:
        \begin{equation}
        \label{1.4}
        {\hat H}=
        \int_{\omega} d^3\! {\mbox{\bf  x}}
        \,
        {\hat e}({\mbox{\bf x}}), 
        \qquad
        {\hat M}=
        \int_{\omega} d^3\! {\mbox{\bf  x}}
        \,
        {\hat \rho}_{m}({\mbox{\bf x}})
        ,
        \qquad
        {\hat \rho}_{m}({\mbox{\bf x}})
        =
        m
        {\hat \psi}^{\scriptscriptstyle\dagger}({\mbox{\bf x}})  
        {\hat \psi}({\mbox{\bf x}})  
        .
        \end{equation}
Obviously it may be uncomfortable to deal with the functions
$u_f({\mbox{\bf x}})$
and to perform discrete sums, even if, but only at a final stage, one can do
approximations like
$
\sum_f h(W_f) u_f({\mbox{\bf x}})
=
\int d\mu({\mbox{\bf  p}}) \,
h({{{\bf p}}^2 \over 2m})
e^{i{
{{\bf p}}\cdot{{\bf x}}
\over
\hbar
}}
$.
The time scale is related to the choice of the relevant
fields in terms of which ${\hat e}({\mbox{\bf x}})$
is given. The picture founded on a ``mass charged'' field
associated to molecules holds if the physics of the system
essentially  depends  on elastic scattering of neutral molecules,
the whole underlying electromagnetic structure being hidden: the
intermolecular (e.g., Lennard-Jones) potential $V(r)$ is a simple
effective representation of the molecular field self-interaction.
A much deeper description of dynamics is possible in
terms of ``electrically charged'' fields (electron and nuclei)
based on QED, but also in this case effective rough elements will
enter in the Hamiltonian density, e.g., the electromagnetic form
factors of nuclei. Unfortunately till now no systematic attempt to base
macrophysics on QED has been developed.
One can expect that the relevance of time scales in
macrophysics, the increasingly deeper descriptions
lowering the time scale, even if at any stage the separation
procedure requires a persistence of some coarse graining of
the dynamical description, indicates a link with the ultraviolet
renormalization problem in field theory: such a link appears
clearer if quantum field theory is seen as the basic theory of
macrosystems, rather than of particles.
Let us now indicate briefly how a piece of macrophysics can be
built based on QSF: hydrodynamics, or
with a slight generalization, kinetic description of a massive
neutral continuum. First of all a classical velocity field is
associated to the continuum and the following basic densities of
conserved quantum observables are considered:
        \begin{eqnarray}
        \label{1.5}
        {\hat e}^{(0)}({\mbox{\bf x}})
        &=&  
        \frac {1}{2m}  
        \left(  
        i\hbar \nabla - m{{\mbox{\bf v}}}
	({\mbox{\bf x}},t)  
        \right)  
        {\hat \psi}^{\scriptscriptstyle\dagger}({\mbox{\bf x}})  
        \cdot  
        \left(  
        -i\hbar \nabla -
        m{{\mbox{\bf v}}}({\mbox{\bf x}},t)  
        \right)  
        {\hat \psi}({\mbox{\bf x}})  
        \nonumber
	\\  
        &\hphantom{=}&  
        +  
        \frac 12  
        \int_{\omega} d^3\! {\mbox{\bf  y}}
	\,  
        {\hat \psi}^{\scriptscriptstyle\dagger}
        ({\mbox{\bf x}})
                {\hat \psi}^{\scriptscriptstyle\dagger}  
        ({\mbox{\bf y}})
        V(|{\mbox{\bf x}}-{\mbox{\bf y}}|)
        {\hat \psi}  
        ({\mbox{\bf y}})
        {\hat \psi}  
        ({\mbox{\bf x}})
        \nonumber
        \\  
        {\hat \rho}^{(0)}_{m}({\mbox{\bf x}})
        &=&  
        {\hat \rho}_{m}({\mbox{\bf x}})
        .
        \end{eqnarray} 
The field
${\hat e}^{(0)}({\mbox{\bf x}})
$
represents the energy density in
the reference frame in which the continuum is locally at rest.
In the kinetic description the mass density
$        {\hat \rho}_{m}({\mbox{\bf x}})
$ should be replaced by the more detailed phase-space distribution
observable
        \[
        {\hat f}({\mbox{\bf x}},{\mbox{\bf p}}) =
        m \sum_{hk}
        {\hat a}_h^{\scriptscriptstyle\dagger}
        \langle
        u_h |   {\hat {\mbox{\sf F}}}^{(1)}
        ({\mbox{\bf x}},{\mbox{\bf p}})  | u_k
        \rangle
        {\hat a}_k                    ,
        \>
        {\hat M} \!= \!
        \int_{\omega} d^3\! {\mbox{\bf x}}
        \int_{R^3} d^3\! {\mbox{\bf p}} \,
        {\hat f}({\mbox{\bf x}},{\mbox{\bf p}})
        \]
constructed on the second-quantized form of the operator density
${\hat {\mbox{\sf F}}}^{(1)}
        ({\mbox{\bf x}},{\mbox{\bf p}})$
on $\omega \times R^3$
  yielding a joint position-momentum observable.
In
correspondence to the velocity field
${{\mbox{\bf v}}}
({\mbox{\bf x}},t)$
and to functions
${ e}^{(0)}({\mbox{\bf x}},t)$,
${ \rho}^{(0)}_{m}({\mbox{\bf x}},t)$
associated at time $t$ to the  operator fields
${\hat e}^{(0)}({\mbox{\bf x}},t)$,
${\hat \rho}^{(0)}_{m}({\mbox{\bf x}},t)$, one
considers the subset of ${\cal K}$  such that:
        \begin{equation}
        \label{1.6}
        e^{(0)}({\mbox{\bf x}},t)
        \!=\!
        {\mbox{\rm Tr}} \!
        \left(  
        {\hat e}^{(0)}({\mbox{\bf x}}) {\hat w}  
        \right)  
         , \,
        {\rho}_{m}({\mbox{\bf x}},t)
        \!=\!
        {\mbox{\rm Tr}}   \!  
        \left(  
        {\hat \rho}_{m}({\mbox{\bf x}}) {\hat w}
        \right)
        ,    \,
        0
        \!=\!
        {\mbox{\rm Tr}}     \!
        \left(  
        {\hat {\mbox{\bf p}}}^{(0)}({\mbox{\bf x}})         {\hat w}
        \right) 
        \end{equation}
with ${\hat {\mbox{\bf p}}}^{(0)}({\mbox{\bf x}})$
the momentum density observable in the reference frame
locally at rest
        \begin{eqnarray*}
        {\hat {\mbox{\bf p}}}^{(0)}({\mbox{\bf x}})
        &=&  
        \frac {1}{2}     \!  
        \left \{  
        \left[
        \left(  
        i\hbar
        \nabla
        - m{{\mbox{\bf v}}}({\mbox{\bf x}},t)
        \right)  
        {\hat \psi}^{\scriptscriptstyle\dagger}({\mbox{\bf x}})  
        \right]  
        {\hat \psi}({\mbox{\bf x}})
        -   \right.
	\\
	&&
	\hphantom{\frac {1}{2}     \!  
        \left \{  \right.
	}
	\left.
        {\hat \psi}^{\scriptscriptstyle\dagger}({\mbox{\bf x}})
        \left(  
        i\hbar \nabla + m{{\mbox{\bf v}}}({\mbox{\bf x}},t)
        \right)  
        {\hat \psi}({\mbox{\bf x}})  
        \right \}  
        \end{eqnarray*}
Then one looks for an element of this subset such that its
von~Neumann entropy $S=-k{\mbox{\rm Tr}
\left(  
{\hat w} \log {\hat w}  
\right)  
}$                   is maximal.
This means a  statistical operator giving the assigned classical
state with highest mixture: i.e., assigning the classical state
has an unmixing role, but no other unmixing process is supposed.
This leads to a
generalized Gibbs
state (Robin, 1990):
        \begin{equation}
        \label{1.7}
        {\hat w}_G (t) \equiv
        {\hat w}[\beta(t) , \mu (t), {{\mbox{\bf v}}}(t)]
        =  
        {  
        e^{-{  
        \int_{\omega} d^3\! {\bf \scriptscriptstyle x} \,  
        \beta({\bf \scriptscriptstyle x},t)  
        \left[  
        {\hat e}^{(0)}({\bf \scriptscriptstyle x})  
        -  
        \mu({\bf \scriptscriptstyle x},t){\hat \rho}_{m}({\bf \scriptscriptstyle x})
        \right]  
        }}  
        \over  
        {\mbox{{\rm Tr}}} \, 
        e^{-{  
        \int_{\omega} d^3\! {\bf \scriptscriptstyle x} \,  
        \beta({\bf \scriptscriptstyle x},t)  
        \left[  
        {\hat e}^{(0)}({\bf \scriptscriptstyle x})  
        -  
        \mu({\bf \scriptscriptstyle x},t){\hat \rho}_{m}({\bf \scriptscriptstyle x})
        \right]  
        }}  
        }\, ,
        \end{equation} 
in the kinetic case ${\hat \rho}_m$ is replaced by ${\hat f}$ and
$\mu$ is a function $\mu({\mbox{\bf x}},{\mbox{\bf p}})$.
The parameters $\beta(t)$, $ \mu (t)$, $ {{\mbox{\bf v}}}(t)$, are
determined by (\ref{1.6}) and will be considered as objective
state variables of the macrosystem;
$S=-k{\mbox{\rm Tr} \,
{\hat w}_G (t)
\log {\hat w}_G (t)
}$
being the  entropy of the  system.
Now the problem arises to make a suitable choice for the representative of the
state at some initial time $t_0$.
According to ``information thermodynamics'' one takes the generalized Gibbs
state determined by the given expectation values at time $t_0$, that is the
most unbiased choice. This approach is certainly satisfying if memory effects
are absent or completely negligible and if no other information about the
system, apart from these expectations, is available, that is to say: the
preparation procedure may be idealized by the istantaneous measurement of the
relevant variables. More general situations, for example memory effects
connected to a macrophysical correlation time, demand a preparation procedure
covering at least the correlation time, thus leading to memory terms in the
representative of the state. The dynamical evolution law must then be fine
enough to keep such effects into account. To circumvent these difficulty
Zubarev, in his definition of the ``non-equilibrium statistical operator''
(Zubarev, 1974), takes the limit $t_0 \rightarrow - \infty$, thus removing any
possible previous memory. This is obtained at the price of introducing a
weighting factor $e^{\varepsilon t}$ that has to be eliminated after the
thermodynamic limit has been taken, thus resorting once more to an infinite
limit. Anyway a suitable memory loss mechanism must be still assumed, typically
decay time of correlation functions.
Our aim is to extract from the dynamics this  memory loss
mechanism, related to a time scale and described inside
the more general framework that we have indicated before.
\par
\section{Time scale and scattering map}
\par
\setcounter{equation}{0}
Let us 
apply the model described in $\S\, 2$ to a dilute gas,
assuming that it has been prepared so that the relevant variables
$
\left \langle
{\hat e}({\mbox{\bf x}})
\right \rangle
$,
$
\left \langle
{\hat \varrho_{\scriptscriptstyle m}}({\mbox{\bf x}})
\right \rangle
$,
$ {\mbox{\bf v}}({\mbox{\bf x}},t)            \left \langle
{\hat \varrho_{\scriptscriptstyle m}}({\mbox{\bf x}})
\right \rangle
=
\left \langle
{\hat {\mbox{\bf p}}}({\mbox{\bf x}})
\right \rangle
$
are smooth enough to provide a macroscopic variation time much
larger than the microscopic collision time $\tau_0$; then, taking
into account the field theoretical structure of
the relevant observables, one has to study expressions of the
form
$
        {\cal U}^{'}
        \left(  
        {\hat a}^{\scriptscriptstyle \dagger}_{h}  
        {\hat a}_{k}
        \right)  
$,
$
        {\cal U}^{'}
        \left(  
        {\hat a}^{\scriptscriptstyle \dagger}_{h_1}  
        {\hat a}^{\scriptscriptstyle \dagger}_{h_2}  
        {\hat a}_{k_2}
        {\hat a}_{k_1}
        \right)
$,
$        {\cal U}^{'}          $ being the time evolution mapping
in Heisenberg picture on ${\cal B}({\cal H})$ (${\cal U}^{'}
        \cdot
        =
        e^{+{{
        i
        \over
         \hbar
        }}{\hat H}t}
        \cdot
        e^{-{{
        i
        \over
         \hbar
        }}{\hat H}t}$),
and to look for an asymptotic representation for $t\gg \tau_0$.
Our procedure essentially consists in transferring to ${\cal
B}({\cal H})$ standard methods of scattering theory related to
${\cal H}$, so the following formulas need no other comment:
        \[
        {\cal H}^{'}_0={i \over \hbar} [{{\hat H}}_0 ,\cdot],
        \quad  
        {\hat H}_0 = \sum_f W_f
        {\hat a}^{\scriptscriptstyle \dagger}_{f}
        {\hat a}_{f},
        \]
         \begin{eqnarray}
         \label{3.2}
         &&
	 {{{\cal U}^{'}(t)}}
         \left(
         {{{\hat a}^{\scriptscriptstyle \dagger}_{h}}{{\hat a}_{k}}}
         \right)
         =
	\left(
         {{{\cal U}^{'}(t)}}{{\hat a}^{\scriptscriptstyle \dagger}_{h}}
         \right)
                \left(
         {{{{\cal U}^{'}(t)}}{{\hat a}_{k}}}
                \right)
	 =
         \\
        && 
	{\int\limits_{-i\infty+\eta}^{+i\infty +  \eta}}{
         dz_1
         \over
             2\pi i
         }     \,     e^{z_1 t}
         \left(
         {
         {1\over{ z_1 - {\cal H}^{'}}}
         {{\hat a}^{\scriptscriptstyle \dagger}_{h}}}
               \right)
         {\int\limits_{-i\infty+\eta}^{+i\infty +  \eta}}{
         dz_2
         \over
             2\pi i
         }       \,   e^{z_2 t}
         \left(
         {
         {1\over{ z_2 - {\cal H}^{'}}}
         {{\hat a}_{k}}
         }
         \right)
         \nonumber
         \end{eqnarray}
        \[
        {{
        {1\over{ z - {\cal H}^{'}}}  
        }}  
        =  
        {{  
        {1\over{ z - {\cal H}^{'}_0}}  
        }} +{{  
        {1\over{ z - {\cal H}^{'}_0}}  
        }}  
        {\cal T}(z){{  
        {1\over{ z - {\cal H}^{'}_0}}  
        }}
	\]  
        \[
	{\cal T}(z)  
        \equiv  
        {\cal V}^{'} + {\cal V}^{'}{{  
        {1\over{ z - {\cal H}^{'}}}  
        }}{\cal V}^{'}      . 
        \]
${\cal T}(z)$, reminiscent of the T-matrix,
plays a central role in this treatment:
it will be called ``scattering map''. Existence of $\tau_0$ means suitable
smoothness properties of ${\cal T}(z)$, so that essentially only
the poles of $({ z - {\cal H}^{'}_0})^{-1}$
contribute to the calculation in (\ref{3.2}), leading to the
representation:
        \begin{equation}
        \label{3.3}
        {{{\cal U}^{'}(t)}}
        \left(  
        {{{\hat a}^{\scriptscriptstyle \dagger}_{h}}{{\hat a}_{k}}}  
        \right)  
        =  
        {{{\hat a}^{\scriptscriptstyle \dagger}_{h}}{{\hat a}_{k}}}  
        + t {\cal L}'
        \left(  
        {{{\hat a}^{\scriptscriptstyle \dagger}_{h}}{{\hat a}_{k}}}
        \right)  
        \qquad
        \tau_0
        \ll
        t
        \ll
        {  
        \hbar  
        \over  
        \left |
        E_h - E_k  
        \right |
        } 
        .
        \end{equation}
Analogous formulas should be written with
$
{\hat a}^{\scriptscriptstyle \dagger}_{h_1}  
{\hat a}^{\scriptscriptstyle \dagger}_{h_2}  
{\hat a}_{k_2}
{\hat a}_{k_1}
$
at place of
$
{\hat a}^{\scriptscriptstyle \dagger}_{h}  
{\hat a}_{k}  
$;
for brevity we skip the derivation of (\ref{3.3}) stressing only the
structural features. ${\cal L}'$ is a linear mapping in ${\cal
B}({\cal H})$ initially defined on the family of linearly
independent elements
$
{\hat a}^{\scriptscriptstyle \dagger}_{h}  
{\hat a}_{k}  
$,
$
{\hat a}^{\scriptscriptstyle \dagger}_{h_1}  
{\hat a}^{\scriptscriptstyle \dagger}_{h_2}  
{\hat a}_{k_2}
{\hat a}_{k_1}
$.
Our approach,
related to relevant field variables in Heisenberg picture,
differs strongly from master equation theory or investigations of
subdynamics (e.g., Prigogine's approach) aiming at a subdynamics
for the statistical operator. The definition of
$ {\cal L}'
\left(
{{{\hat a}^{\scriptscriptstyle \dagger}_{h}}{{\hat
a}_{k}}}
\right) $,  operator in the Fock-space of QSF, is at first sight
very simple:
        \begin{eqnarray*}
	&&
	        \!\! \!\!  \!\!    \!\! \!
	{\cal L}^{'}
        \left(
        {\hat a}^{\scriptscriptstyle \dagger}_{h}
        {\hat a}_{k} 
        \right)
        =
	\\
        &&
	\!\! \!\!  \!\!  \!\!  \!
       	=
        {i\over\hbar}  
        \left[  
        {\hat H}_{\rm \scriptscriptstyle eff},
        {\hat a}^{\scriptscriptstyle \dagger}_{h}  
        {\hat a}_{k}  
        \right]  
        - {1\over \hbar}  
        \left(  
        \left[  
        {\hat \Gamma}^{(2)} , {\hat a}^{\scriptscriptstyle\dagger}_h
        \right]  
        {\hat a}_k  
        -  
        {\hat a}^{\scriptscriptstyle\dagger}_h  
        \left[  
        {\hat \Gamma}^{(2)}, {\hat a}_k  
        \right]  
        \right)  
        +  
        {1\over\hbar} \sum_\lambda  
        {\hat R}^{(2)}_{h \lambda}{}^{\dagger}  
        {\hat R}^{(2)}_{k \lambda}
        \end{eqnarray*}        
	\begin{eqnarray}
        \label{3.4}
        {\hat H}_{\rm \scriptscriptstyle eff}
        &=&  
        \sum_f W_f
        {\hat a}_f^{\scriptscriptstyle\dagger} {\hat a}_f  
        + {1\over 2}  
        \sum_{l_1 l_2 \atop f_1 f_2}  
        {\hat a}_{l_1}^{\scriptscriptstyle\dagger}
        {\hat a}_{l_2}^{\scriptscriptstyle\dagger}  
        V{}_{l_1 l_2 f_2 f_1}^{eff}
        {\hat a}_{f_2}  
        {\hat a}_{f_1}  
        \nonumber \\
        {\hat \Gamma}^{(2)}
        &=&
        {1\over4}
         \sum_{h \lambda}  
        {\hat R}^{(2)}_{h \lambda}{}^{\dagger}  
        {\hat R}^{(2)}_{h \lambda} 
        \qquad
        {\hat R}^{(2)}_{k \lambda}
        =
        \sum_{f_1 f_2}  
        {R}_{k \lambda f_2 f_1}
        {\hat a}_{f_2}  
        {\hat a}_{f_1}        ;
        \end{eqnarray}
the coefficients
$V{}_{l_1 l_2 f_2 f_1}^{eff}
$,
${R}_{k \lambda f_2 f_1}$ are directly related with the two-particle
T-matrix for  scattering produced by the potential $V(r)$
appearing in (\ref{1.1}). This simple structure comes from a ``one
interacting mode approximation'', appropriate for
a not too dense system, by which only two-particle collisions are considered.
However the
definition of ${\cal L}'$ is more
tricky                           since by quantum non-separability
Pauli principle corrections must arise. In fact the coefficients
$V{}_{l_1 l_2 f_2 f_1}^{eff}
$,
${R}_{k \lambda f_2 f_1}$ are 
not c-numbers, but are operator valued in the Fock-space
of QSF, diagonal in the basis created by
$        {\hat a}_{f}^{\scriptscriptstyle\dagger} $; to
transform an element $|\ldots { n}_f \ldots \rangle$ of this
basis by the operator
$ {\cal L}'
\left(
{{{\hat a}^{\scriptscriptstyle \dagger}_{h}}{{\hat
a}_{k}}}
\right) $
one applies to it    
the r.h.s. of
(\ref{3.4}) where
the coefficients
$V{}_{l_1 l_2 f_2 f_1}^{eff}
$,
${R}_{k \lambda f_2 f_1}$ are 
functionals of the configuration $
\left \{
n_f
\right \}
$.
${\cal L}'$ as given by (\ref{3.4})
generates a positive dynamics, i.e. $({\cal
I}+\tau {\cal L}')$ is positive at first order in $\tau$, for
$\tau> 0$. Actually one has a stronger property:
        \begin{equation}
        \label{3.5}
        \sum_{hk}
        \langle  
        \psi_h  
        |
        \left[
        \left(
        {\cal I}+\tau {\cal L}'
        \right)
        \left(
        {\hat a}^{\scriptscriptstyle \dagger}_{h}
        {\hat a}_{k}
        \right)
        \right]
        \psi_k  
        \rangle  
        >0  
        \qquad 
        \forall \ 
        \left \{
        \psi_k
        \right \}           ,\
        \psi_k \in {\cal H} ;
        \end{equation}
due to the fact that (\ref{3.5}) holds  for $\tau>0$ only,
irreversibility is introduced. Property (\ref{3.5}) looks as a
straightforward adjustment to Fock-space structure of the
well-known complete positivity notion for a  mapping   ${\cal M}$
 on ${\cal B}({\cal H})$:
        \[
        \sum_{hk}
        \langle
        \psi_h
        |
        {\cal M}
        \left(
        {\hat A}^{\scriptscriptstyle \dagger}_{h}
        {\hat A}_{k}
        \right)
        \psi_k
        \rangle
        >0
        \quad
        \forall\ 
        \left \{
         \psi_k
        \right \}           , \
        \left \{
         {\hat A}_k
        \right \}           ,\
        \psi_k \in {\cal H} , {\hat A}_k \in {\cal B}({\cal H})
        .
        \]
The link between
${\hat \Gamma}^{(2)}$            and ${\hat R}^{(2)}_{k \lambda}$
implies mass conservation, ${\cal L}' {\hat M}=0$, and
one expects that also energy conservation, ${\cal L}' {\hat E}=0
$,
exactly holds, but some further analysis is necessary. Now the
following assumption becomes very natural: the generalized Gibbs
states related to the relevant observables
${\hat e}({\mbox{\bf x}})$,
${\hat \rho}_{m}({\mbox{\bf x}})$
can also be used to obtain the expectations of the ``coarse
grained'' time derivatives of these variables, i.e., ${\hat w}$
and ${\cal L}'$ are respectively states and evolution map tuned
to the time scale $t\gg \tau_0$.
Then by  eq.~(\ref{1.6}) one has the evolution equation for
the generalized Gibbs states:
        \begin{equation}
        \label{3.7}
        {  
        d  
        \over  
         dt  
        }  
        {\mbox{\rm Tr}}  
        \left(  
        {\hat A}  
        {\hat w}[\beta(t) , \mu (t), {{\mbox{\bf v}}}(t)]  
        \right)  
        =  
        {\mbox{\rm Tr}}  
        \left(  
        ({\cal L}^{'}  
        {\hat A})  
        {\hat w}[\beta(t) , \mu (t), {{\mbox{\bf v}}}(t)]  
        \right)  ,
        \end{equation} 
where $
{\hat A}=                                   {{\hat e}}^{(0)}({\mbox{\bf x}}),
{  {\hat \rho}}^{(0)}_{m}({\mbox{\bf x}}),
{  {\hat {\mbox{\bf p}}}}^{(0)}({\mbox{\bf x}})
$,
thus providing
a set of closed evolution equations for the objective state
parameters
$\beta(t)$, $ \mu (t)$, $ {{\mbox{\bf v}}}(t)$. Choosing
$        {\hat A}
        =
        {\hat a^{\scriptscriptstyle \dagger}_{h}}  
        {\hat a_{k}} $, by inspection of the r.h.s. of eq.~(\ref{3.4})
one can immediately recognize the relationship with
Boltzmann equation: the last two terms of (\ref{3.4}) have
the typical form of a collision  operator,
the
$        {\hat \Gamma}^{(2)} ,
        {\hat R}^{(2)}_{h \lambda}{}^{\dagger} \cdot
        {\hat R}^{(2)}_{k \lambda}
 $
contributions being respectively the loss and the gain part.
This description
avoids any
factorization of many-particle distribution functions. The
dynamics on the coarse grained time scale $t\gg \tau_0$
looses any memory of previous states
and is described by the irreversible map ${\cal L}'$.
If the
approximation leading to ${\cal L}'$ does not work, one
expects that memory effects can appear and that the starting
point could be shifted from (\ref{1.1}) to the QED
Hamiltonian.            
To conclude these considerations about an
isolated   macrosystem  described by a  statistical operator
$       {{\hat \varrho}_{\rm M}}(t)
$ and an Hamiltonian ${{\hat H}}_{\rm M}$,
let us show how the simplest breaking of the isolation of
this system leads to the concept of a
microsystem. Consider
a Hamiltonian
${\hat H}$ and a statistical operator ${\hat \varrho}(t)$
of the form:
        \begin{equation}
        \label{3.8}
        {{\hat H}}={{\hat H}}_0 + {{\hat H}}_{\rm M} +
        {{\hat V}}  \qquad
        {{\hat H}}_0 = \sum_p
        {E_p}
        {{\hat b}^{\scriptscriptstyle \dagger}_{q}}
        {{\hat b}_{p}}
         \qquad
        \left[{{{\hat b}_{{p}}},{{\hat b}^{\scriptscriptstyle
        \dagger}_{q}}}\right]_{\mp}=\delta_{pq} ;
        \end{equation}
        \begin{equation}
        \label{3.9}
        {\hat \varrho}(t)=  
        \sum_{{q} {p}}{}
        {{\hat b}^{\scriptscriptstyle \dagger}_{q}}
        {{\hat \varrho}_{\rm M}}(t)
        {{\hat b}_{{p}}}
        {{ \varrho}}_{qp}(t)
                \qquad
        {{\hat b}_{{p}}}{{\hat \varrho}_{\rm M}}=0    .
        \end{equation}
Due to the condition ${{\hat b}_{{p}}}{{\hat \varrho}_{\rm
M}}=0$, the QSF
${\hat \phi} ({\mbox{\bf x}})
=
\sum_q {\tilde u}_q ({\mbox{\bf x}}) {\hat
b}_q $, is either related to a different particle or
to other modes ${\tilde u}_g$ (e.g., confinement in
$ \tilde\omega \supset \omega$) than those involved
in (\ref{1.2}).
Indicating with $
{\hat Q}=\sum_q {{\hat b}^{\scriptscriptstyle \dagger}_{q}}
        {{\hat b}_{{q}}}
$                   the related charge, one has
$
{\hat Q}{{\hat \varrho}_{\rm
M}}=0
$, $
{\hat Q}{{\hat \varrho}}={\hat \varrho}
$: this indicates the ``elementary'' nature of the
change of ${\hat \varrho}$.
Under suitable
conditions, the new relevant variables
$        {\hat A}
= \sum_{h,k}
{\hat b^{\scriptscriptstyle \dagger}_{h}}
{A}_{hk}
{\hat b_{k}} $ can be treated by the same
procedure we have indicated
before.
Using the reduction formula:
        \[
        {\hbox{\rm Tr}}_{{\cal H}}
        \left(  
        {{\hat A}{\hat \varrho}}(t)  
        \right)  
        =  
        {\hbox{\rm Tr}}_{{{\cal H}^{(1)}}}  
        \left(  
        {\hat {{\mbox{\sf A}}}}^{(1)} {{\hat \varrho}}^{(1)}(t)
        \right)     
        \]
	\[
        {\hat {{\mbox{\sf A}}}}^{(1)}
        =
        \sum_{hk}  |u_h \rangle A_{hk} \langle u_k |
        \quad
        {\hat \varrho}^{(1)}=
        \sum_{qp}  |u_q \rangle \varrho_{qp} \langle u_p |,
        \]
${\cal H}^{(1)}$ being the Hilbert space spanned by ${\tilde
u}_q$,
the reduced dynamics
can be interpreted as a
microsystem described in the Hilbert space ${\cal H}^{(1)}$,
with observables                    ${\hat {{\mbox{\sf
A}}}}^{(1)}$ and preparation ${\hat \varrho}^{(1)}$. For
${\hat \varrho}^{(1)}$ a master equation is found (Lanz
and Vacchini, 1997a, 1997b),
describing both the optical behavior associated to the
analogous of
${\hat H}_{\rm \scriptscriptstyle eff}$ in (\ref{3.4}) and a
incoherent part related to the other part of ${\cal L}'$.
The role of the first part is enhanced in the typical setup
of particle interferometry and in this way one comes back to
the one particle Schr\"odinger equation;
the second part
describes Brownian motion and thermalization of the particle
inside matter.
We recall that an objective reinterpretation
of the dynamics of the new variables due to the
non-Hamiltonian evolution is possible in terms of a
statistical description of trajectories of the non-isolated
particle (Lanz and Melsheimer, 1993), however a systematic extension of this
objectifying procedure to the relevant macroscopic
variables considered in $\S \, 1$, for which in the present
treatment only the expectation values have been considered,
is an open question. The statistical operator  (\ref{3.9})
describes a microsystem + a  macrosystem  without a reaction
of the microsystem on the macrosystem; so this is not yet
enough to treat  in
the context of the theory of macrosystems the typical setting: 
source -- detector.
\vskip 15pt
{\parindent = 0 pt
{\LARGE References}
\vskip 10pt

{Davies, E.~B.}
(1976).
{\it Quantum Theory of Open
Systems}, {Academic Press}, {London}.

Holevo, A.~S.
(1982).
{\it Probabilistic and Statistical Aspects of
Quantum Theory}, {North Holland}, {Amsterdam}{}.

Kraus, K.
(1983).
States, Effects and Operations, in {Lecture Notes in
Physics}, Volume 190, {Springer}, {Berlin}.

Lanz, L., and Melsheimer, O.
(1993). 
Quantum Mechanics and Trajectories,
in
{\it Symposium On the Foundations of Modern Physics},
Busch, P., Lahti, P.~J., and Mittelstaedt, P., eds., World
Scientific, Singapore, p.233-241.

{Lanz, L., and Vacchini, B.}
(1997a).
{\it International Journal of Theoretical Physics},
{\bf 36},
{67}.

{Lanz, L., and Vacchini, B.}
(1997b).
{\it Physical Review A},
{\bf 56},
{4826}.

{Ludwig, G.}
(1983).
{\it Foundations of Quantum
Mechanics}, {Springer}, {Berlin}.

{Robin, W.~A.}
(1990).
{\it Journal of Physics A},
{\bf 23},  
{2065}.
  
{Zubarev, D.~N.}
(1974).
{\it Non-equilibrium statistical thermodynamics},
Consultant Bureau, New York.
}
\end{document}